\documentclass[preprint,numbers,5p]{elsarticle}

\usepackage{graphicx}

\usepackage{amssymb}

\usepackage{amsmath}

\biboptions{sort&compress}

\journal{Physics Letters A}

\usepackage{color} 
\usepackage{url}
\usepackage[latin1]{inputenc}
\usepackage[T1]{fontenc}

\begin{document}

\begin{frontmatter}



\title{Comment on ``Convergence towards asymptotic state in 1-D mappings: A scaling investigation''}


\author[endUSP,email]{Mauricio Girardi-Schappo}
\author[endUFSC]{Marcelo H. R. Tragtenberg}
\address[endUSP]{Departamento de F\'isica, FFCLRP, Universidade de S\~ao Paulo, 14040-901, Ribeir\~ao Preto, S\~ao Paulo, Brazil}
\address[endUFSC]{Departamento de F\'isica, Universidade Federal de Santa Catarina, 88040-900, Florian\'opolis, Santa Catarina, Brazil}
\address[email]{girardi.s@gmail.com}

\begin{abstract}
Nonequilibrium phase transitions are characterized by the so-called critical exponents,
each of which is related to a different observable. Systems that share the same set of
values for these exponents also share the same universality class. Thus, it is important
that the exponents are named and treated in a standardized framework. In this comment,
we reinterpret the exponents obtained in [Phys Lett A 379:1246-12 (2015)]
for the logistic and cubic maps in order to correctly state the universality class of their
bifurcations, since these maps may describe the mean-field solution of stochastic spreading processes.
\\$ $
\\
\textbf{DOI:} 10.1016/j.physleta.2019.126031
\end{abstract}


\end{frontmatter}


\citet{teixeira2015} studied the generalized logistic map,
\begin{equation}
\label{eq:logmap}
x_{t+1}=R x_t(1-x_t^a)\:,
\end{equation}
where $R$ and $a=1$ (the usual logistic map~\cite{may1976}) or $a=2$ (the cubic map) are control parameters,
and $t$ is a discretized time step. The variable $x_t\in[0;1]$ usually represents a generic population
density that may linearly increase or nonlinearly decrease over time, depending on the values
of $R$ and $a$~\cite{may1976}.
The asymptotic stationary solution of equation~\eqref{eq:logmap} (also known as fixed point -- FP)
is found by imposing $x_{t+1}=x_t=x^*$~\cite{may1976}. The FP's of the generalized logistic map,
other than the trivial solution $x^*=0$ for $R<R_c\equiv1$, are:
\begin{align}
\label{eq:fp1}
x^*&=\dfrac{R-1}{R}           &(\textnormal{for }a=1; R>R_c)\\
\label{eq:fp2}
x^*&=\pm\sqrt{\dfrac{R-1}{R}} &(\textnormal{for }a=2; R>R_c)
\end{align}

The authors did a very nice job characterizing the scaling laws governing the convergence time to the
FP of the maps given in equation~\eqref{eq:logmap}. They calculated what they called
\textit{critical exponents} for the transcritical ($a=1$) and pitchfork ($a=2$)
bifurcations.
The calculations done in~\cite{teixeira2015} are completely sound and well supported by mathematical
and numerical evidence collected by the authors.
However, the term \textit{critical exponents} has a special meaning in the Nonequilibrium Phase
Transitions literature because they are used to define universality
classes~\cite{odorReview2004,noneqPhaseTrans2008}.
These classes are usually dependent on the collective behavior of systems, its symmetries, its dimensionality,
and so on~\cite{odorReview2004}, and are fundamental for a coherent theory of phase transitions in
nonequilibrium systems.
There has been an effort in the Phase Transitions community to standardize the
nomenclature and definition of critical exponents, assigning each individual exponent to a unique Greek letter.
Conversely, each exponent is uniquely linked to an observable of the system.

Recently, it was shown that the logistic map describes the population density
of the mean-field phase transition between the inactive ($x^*=0$)
and active ($x^*>0$) states of a spreading process (see equation (19) in~\cite{brochiniPhasTrans2016} and equation (4) in~\cite{girardiSOCBal2019}),
and that the critical point at $R=R_c=1$ pertains to the mean-field
directed percolation (DP) universality
class~\cite{brochiniPhasTrans2016,girardiSOCBal2019}.
This development puts the logistic map in the context of the well-known absorbing state phase
transitions,
and is not consistent with the definition
of the exponents $\alpha$, $\beta$, $z$ and $\delta$ done by \citet{teixeira2015}.
Therefore, we revisit the definition of the critical exponents of \citet{teixeira2015}
to match the long standing standard nomenclature in the literature.
From now on, we will use $\alpha_1$, $\beta_1$, $z_1$ and $\delta_1$
for the critical exponents defined in~\cite{teixeira2015} in order
to correctly distinguish them from the usual nonequilibrium exponents,
$\alpha$ (the exponent of the order parameter \textit{vs.} time), $\beta$ (the exponent of the 
order parameter \textit{vs.} a temperature-like parameter), $\delta$ (the survival probability
exponent), and $z$ (the dynamical exponent)~\cite{noneqPhaseTrans2008}.

The exponent $\beta_1$ [equation (3) in~\cite{teixeira2015}] is,
in fact, the standard exponent $\alpha$ describing the decay of the
average activity towards the stationary state $x^*=0$ on the bifurcation point
$R=R_c$~\cite[Table 4.1]{noneqPhaseTrans2008}:
\begin{equation}
\label{eq:alpha}
x_t\sim t^{-\alpha}\:.
\end{equation}
Thus, the values given by~\citet{teixeira2015} for $\beta_1$ are the values of $\alpha$,
\textit{i.e.} $\alpha=1$ for the logistic map ($a=1$) and $\alpha=1/2$ for the cubic
map ($a=2$)~\cite{teixeira2015}.

The standard exponent $\beta$ comes from the stationary state of the order parameter
(given by the population density, $x^*$) evaluated very close to the critical (\textit{i.e.}, the
bifurcation) point, $R_c=1$~\cite[Table 4.1]{noneqPhaseTrans2008}:
\begin{equation}
\label{eq:beta}
x^*\sim (R-R_c)^{\beta}\:.
\end{equation}
None of the exponents calculated by~\citet{teixeira2015} corresponds
to the $\beta$ given in equation~\eqref{eq:beta}.
We can expand to first order the FP equations~\eqref{eq:fp1} and~\eqref{eq:fp2} for $R\gtrsim1$
to obtain $x^*\sim(R-1)^{\beta}$, with $\beta=1$ for the logistic map
and $\beta=1/2$ for the cubic map. The apparent match between the values of $\beta_1$
and $\beta$ comes from the fact the the logistic equation describes the population density at a
mean-field level~\cite{brochiniPhasTrans2016,girardiSOCBal2019}. For
other dimensionalities, the values of $\beta_1$ (\textit{i.e.}, the standard $\alpha$)
and $\beta$ do not coincide~\cite[Table 4.3]{noneqPhaseTrans2008}.

\citet{teixeira2015} defined the exponent $\delta_1$ [equation (10) in their paper]
as the decay exponent of the autocorrelation length, $\xi_{\parallel}$, of $x_t$ as a function
of the bifurcation parameter $(R-R_c)$.
This is the precise definition of the standard exponent $\nu_{\parallel}$, via
the scaling law~\cite[Table 4.1]{noneqPhaseTrans2008}:
\begin{equation}
\label{eq:xi}
\xi_{\parallel}\sim|R-R_c|^{-\nu_{\parallel}}\:,
\end{equation}
such that $\delta_1=-\nu_{\parallel}$.
Thus, the values estimated numerically and analytically by the authors, $\delta_1=-1$,
for both the logistic and the cubic map yield $\nu_{\parallel}=1$.

We can assume that the crossover time calculated by the authors, $n_x$ [equation (4)
in~\cite{teixeira2015} defining the $z_1$ exponent],
corresponds to $\xi_{\parallel}$, which decays algebraically with $t$ when $R=R_c=1$
as~\cite[p.107]{noneqPhaseTrans2008}
\begin{equation}
\label{eq:xi2}
\xi_{\parallel}\sim t^{-\beta/\nu_{\parallel}}\:.
\end{equation}
This assumption also requires that the time, $T$, spent by the logistic map lurking on the
initial condition, $x_0$, before
decaying is $T\propto x_0$ -- in fact, this lead \citet{teixeira2015} to define
their equation (4).
Thus, from Eq.~\eqref{eq:xi2}, $z_1$ may be expressed in terms of the standard exponents,
$z_1=-\nu_{\parallel}/\beta$. The authors fitted the values $z_1=-1.0002(3)$ and
$z_1=-2.001(2)$ for the logistic and the cubic map, respectively. These values agree with
the ratio between the standard exponents $\beta$ [from Eq.~\eqref{eq:beta}]
and $\nu_{\parallel}$ (\textit{i.e.}, the author's $-\delta_1$),
yielding $z_1=-1/1=-1$ for the logistic map, and $z_1=-1/(1/2)=-2$
for the cubic map.

The exponent $\alpha_1$ defined by \citet{teixeira2015} has
no counterpart in the standard literature of nonequilibrium phase 
transitions~\cite{odorReview2004,noneqPhaseTrans2008}.
However, notice that the definition, $x_t\propto x_0^{\alpha_1}$
[equation (2) in~\cite{teixeira2015}],
determines the expansion rate over time of the initial condition, $x_0$.
It is known that the Lyapunov exponent for both the transcritical and the pitchfork
bifurcations is zero~\cite{bakerChaos}. Equation (21) in~\cite{teixeira2015} can be
transformed into the definition of the Lyapunov exponent, $\lambda_L$,
\begin{equation}
\label{eq:lyap}
|\Delta x_t|=|\Delta x_0|\exp{(\lambda_L t)}\:,
\end{equation}
where Eq.~\eqref{eq:lyap} describes how a variation in the initial condition,
$|\Delta x_0|$, evolves with time.
Thus, from equation (21) in~\cite{teixeira2015}
and our Eq.~\eqref{eq:lyap}, $\lambda_L=R-1=0$ for $R=1$ (\textit{i.e.},
on the bifurcation point). A null Lyapunov exponent means that
the initial condition does not expand nor shrink, causing the
observed value $\alpha_1=1$.

The set of exponents $\beta=1$, $\alpha=1$ and $\nu_{\parallel}=1$,
together with $\gamma=1$ and $\delta_h=2$
from~\cite{brochiniPhasTrans2016,girardiSOCBal2019},
puts the transcritical bifurcation of the FP of the logistic map
into the mean-field DP universality class. For the cubic map,
the set of values $\beta=1/2$, $\alpha=1/2$ and $\nu_{\parallel}=1$,
together with its cubic nonlinearity
[compare it with equation~(5.14) in~\cite{noneqPhaseTrans2008} for $g\rightarrow0$],
puts its pitchfork bifurcation in the mean-field
tricritical directed percolation (TDP) universality class~\cite[Table 5.2]{noneqPhaseTrans2008}.

Bifurcations and nonequilibrium phase transitions share similar relations, and 
the context in which the model is studied evokes either one or the other
as the background theory to explain the observed changes of macroscopic behavior
occurring in the system. Nevertheless, a unification of both approaches could only benefit
both fields. To that extent, a standard shared nomenclature is essential to correctly
define the critical exponents and bring together the two frameworks.
The calculations done by~\citet{teixeira2015} are
correct. However, we reinterpreted their exponents in the light of absorbing phase
transitions and explained the apparent nonuniversal behavior of the $\beta_1$ exponent
(\textit{i.e.}, the standard $\alpha$) observed by the authors as $a$ changes from 1
(the logistic map) to 2 (the cubic map).
The apparent nonuniversality happened because the logistic and cubic maps pertain to
different mean-field universality classes: the DP and the TDP, respectively.

\section*{Acknowledgments}
Article produced through the
FAPESP Research, Innovation and Dissemination
Center for Neuromathematics.
M.G.-S. acknowledges financial support from Grant No. 2018/09150-9, S. Paulo Research Foundation.

\bibliographystyle{model6-num-names}

\begin{thebibliography}{7}
\providecommand{\natexlab}[1]{#1}
\providecommand{\url}[1]{\texttt{#1}}
\providecommand{\urlprefix}{URL }
\expandafter\ifx\csname urlstyle\endcsname\relax
  \providecommand{\doi}[1]{doi:\discretionary{}{}{}#1}\else
  \providecommand{\doi}{doi:\discretionary{}{}{}\begingroup
  \urlstyle{rm}\Url}\fi
\providecommand{\eprint}[2][]{\url{#2}}
\providecommand{\BIBand}{and}
\providecommand{\bibinfo}[2]{#2}
\ifx\xfnm\undefined \def\xfnm[#1]{\unskip,\space#1}\fi
\makeatletter\def\@biblabel#1{#1.}\makeatother
\bibitem[{Teixeira et~al.(2015)Teixeira, Rando, Geraldo, Costa{ }Filho, de{
  }Oliveira and Leonel}]{teixeira2015}
\bibinfo{author}{Teixeira\xfnm[ R.M.N.]}, \bibinfo{author}{Rando\xfnm[ D.S.]},
  \bibinfo{author}{Geraldo\xfnm[ F.C.]}, \bibinfo{author}{Costa{ }Filho\xfnm[
  R.N.]}, \bibinfo{author}{de{ }Oliveira\xfnm[ J.A.]},
  \bibinfo{author}{Leonel\xfnm[ E.D.]}.
\newblock \bibinfo{title}{Convergence towards asymptotic state in 1-{D}
  mappings: {A} scaling investigation}.
\newblock \emph{\bibinfo{journal}{Phys Lett A}}
  \bibinfo{year}{2015};\bibinfo{volume}{379}:\bibinfo{pages}{1246--1250}.
\newblock \doi{\bibinfo{doi}{10.1016/j.physleta.2015.02.019}}.
\bibitem[{May(1976)}]{may1976}
\bibinfo{author}{May\xfnm[ R.M.]}.
\newblock \bibinfo{title}{Simple mathematical models with very complicated
  dynamics}.
\newblock \emph{\bibinfo{journal}{Nature}}
  \bibinfo{year}{1976};\bibinfo{volume}{261}:\bibinfo{pages}{459--467}.
\bibitem[{{\'O}dor(2004)}]{odorReview2004}
\bibinfo{author}{{\'O}dor\xfnm[ G.]}.
\newblock \bibinfo{title}{Universality classes in nonequilibrium lattice
  systems}.
\newblock \emph{\bibinfo{journal}{Rev Mod Phys}}
  \bibinfo{year}{2004};\bibinfo{volume}{76(3)}:\bibinfo{pages}{663--724}.
\bibitem[{Henkel et~al.(2008)Henkel, Hinrichsen and
  L{\"u}beck}]{noneqPhaseTrans2008}
\bibinfo{author}{Henkel\xfnm[ M.]}, \bibinfo{author}{Hinrichsen\xfnm[ H.]},
  \bibinfo{author}{L{\"u}beck\xfnm[ S.]}.
\newblock \bibinfo{title}{Non-Equilibrium Phase Transitions}.
\newblock \bibinfo{address}{Dordrecht, The Netherlands}:
  \bibinfo{publisher}{Springer}; \bibinfo{year}{2008}.
\bibitem[{Brochini et~al.(2016)Brochini, Costa, Abadi, Roque, Stolfi and
  Kinouchi}]{brochiniPhasTrans2016}
\bibinfo{author}{Brochini\xfnm[ L.]}, \bibinfo{author}{Costa\xfnm[ A.A.]},
  \bibinfo{author}{Abadi\xfnm[ M.]}, \bibinfo{author}{Roque\xfnm[ A.C.]},
  \bibinfo{author}{Stolfi\xfnm[ J.]}, \bibinfo{author}{Kinouchi\xfnm[ O.]}.
\newblock \bibinfo{title}{Phase transitions and selforganized criticality in
  networks of stochastic spiking neurons}.
\newblock \emph{\bibinfo{journal}{Sci Rep}}
  \bibinfo{year}{2016};\bibinfo{volume}{6}:\bibinfo{pages}{35831}.
\newblock \doi{\bibinfo{doi}{10.1038/srep35831}}.
\newblock \urlprefix\url{http://dx.doi.org/10.1038/srep35831}.
\bibitem[{Girardi{-}{S}chappo et~al.(2019)Girardi{-}{S}chappo, Brochini, Costa,
  Carvalho and Kinouchi}]{girardiSOCBal2019}
\bibinfo{author}{Girardi{-}{S}chappo\xfnm[ M.]},
  \bibinfo{author}{Brochini\xfnm[ L.]}, \bibinfo{author}{Costa\xfnm[ A.A.]},
  \bibinfo{author}{Carvalho\xfnm[ T.T.A.]}, \bibinfo{author}{Kinouchi\xfnm[
  O.]}.
\newblock \bibinfo{title}{Self-organized critical balanced networks: a unified
  framework}.
\newblock \emph{\bibinfo{journal}{arXiv}}
  \bibinfo{year}{2019};:\bibinfo{pages}{1906.05624 [nlin.AO]}.
\bibitem[{Baker and Gollub(1996)}]{bakerChaos}
\bibinfo{author}{Baker\xfnm[ G.L.]}, \bibinfo{author}{Gollub\xfnm[ J.P.]}.
\newblock \bibinfo{title}{Chaotic Dynamics: An Introduction}.
\newblock \bibinfo{address}{Cambridge, UK}: \bibinfo{publisher}{Cambridge
  University Press}; \bibinfo{year}{1996}.

\end{thebibliography}
 \newcommand{\noop}[1]{}







\end{document}